\documentclass[conference]{IEEEtran}
\IEEEoverridecommandlockouts
\usepackage{cite}
\usepackage{amsmath,amssymb,amsfonts}
\usepackage{algorithmic}
\usepackage{graphicx}
\usepackage{xspace}
\usepackage{textcomp}
\usepackage{url}   
\usepackage{xcolor}
\def\BibTeX{{\rm B\kern-.05em{\sc i\kern-.025em b}\kern-.08em
    T\kern-.1667em\lower.7ex\hbox{E}\kern-.125emX}}

\begin{document}

\newcommand{\ttt}[1]{\texttt{\small{#1}}}
\newcommand{\attackI}{\mbox{$\mathcal{A}_1$}\xspace}
\newcommand{\attackII}{\mbox{$\mathcal{A}_2$}\xspace}
\newcommand{\attackIII}{\mbox{$\mathcal{A}_3$}\xspace}
\newcommand{\attackIV}{\mbox{$\mathcal{A}_4$}\xspace}
\newcommand{\attackV}{\mbox{$\mathcal{A}_5$}\xspace}
\newcommand{\attackVI}{\mbox{$\mathcal{A}_6$}\xspace}
\newcommand{\attackVII}{\mbox{$\mathcal{A}_7$}\xspace}
\newcommand{\msI}{\mbox{$\mathcal{M}_1$}\xspace}
\newcommand{\msII}{\mbox{$\mathcal{M}_2$}\xspace}
\newcommand{\msIII}{\mbox{$\mathcal{M}_3$}\xspace}
\newcommand{\msIV}{\mbox{$\mathcal{M}_4$}\xspace}
\newcommand{\msV}{\mbox{$\mathcal{M}_5$}\xspace}
\newcommand{\msVI}{\mbox{$\mathcal{M}_6$}\xspace}
\newcommand{\msVII}{\mbox{$\mathcal{M}_7$}\xspace}
\newcommand{\head}[1]{\textnormal{\textbf{#1}}}

\title{Et tu, Blockchain? Outsmarting Smart Contracts via Social Engineering}

\author{\IEEEauthorblockN{Nikolay Ivanov}
\IEEEauthorblockA{\textit{Dept. of Computer Science and Engineering} \\
\textit{Michigan State University}\\
East Lansing, MI, USA \\
ivanovn1@msu.edu}
\and
\IEEEauthorblockN{Qiben Yan}
\IEEEauthorblockA{\textit{Dept. of Computer Science and Engineering} \\
\textit{Michigan State University}\\
East Lansing, MI, USA \\
qyan@msu.edu}
}

\maketitle

\begin{abstract}
We reveal six zero-day social engineering attacks in Ethereum, and subdivide them into two classes: Address Manipulation and Homograph. We demonstrate the attacks by embedding them in source codes of five popular smart contracts with combined market capitalization of over \$29 billion, and show that the attacks have the ability to remain dormant during the testing phase and activate only after production deployment. We analyze 85,656 open source smart contracts and find 1,027 contracts that can be directly used for performing social engineering attacks. For responsible disclosure, we contact seven smart contract security firms. In the spirit of open research, we make the source codes of the attack benchmark, tools, and datasets available to the public\footnote{\url{https://nick-ivanov.github.io/se-info/}}.
\end{abstract}

\begin{IEEEkeywords}
Ethereum, social engineering, security, smart contracts
\end{IEEEkeywords}


\section{Smart Contract Security}
The blockchain technology is ramping up its popularity and market capitalization.
A major reason of its vast expansion is the ability to support \emph{smart contracts} --- decentralized programs that can enforce execution of protocols without any third party or mutual trust.
As of December 2020, the Tether USD smart contract had more than 2.1 million users with about \$36 billion in daily transaction volume~\cite{etherscan-tokens}.
Unfortunately, smart contracts suffer from vulnerabilities, manifested by recent hacks with multimillion-dollar damages~\cite{mehar2019understanding,palladino2017parity}.
A recent analysis by Zhou et al. reveals an ongoing evolution of attacks in smart contracts~\cite{zhou2020ever}.
To avoid devastating consequences of smart contract hacks, a number of security auditing tools have been developed to detect smart contract vulnerabilities~\cite{tsankov2018securify,luu2016making}, such as reentrancy, integer overflow, etc., most of which are smart contract code vulnerabilities. However, smart contracts are designed and implemented by human developers to interact with human users, in which the human is the central component of a smart contract ecosystem~\cite{ivanov2021rectifying}. Yet, most existing smart contract security studies do not take the human factor into account.

\section{Social Engineering}
We show that the Ethereum platform enables \emph{social engineering attacks}. Social engineering attacks have been known across multiple technologies, such as landline phones and industrial networks. When existing software and hardware defense reduces the attack surface, the adversaries resort to exploiting human cognitive bias --- the weakest link in many security systems. This poster is the summary of our full paper ``Targeting the Weakest Link: Social Engineering Attacks in Ethereum Smart Contracts''~\cite{ivanov2021targeting}, which presents the first investigation of the possibility, vectors, and impact of social engineering attacks in smart contracts, as well as defense against these attacks.

\begin{figure}
    \centering
    \includegraphics[width=0.65\linewidth]{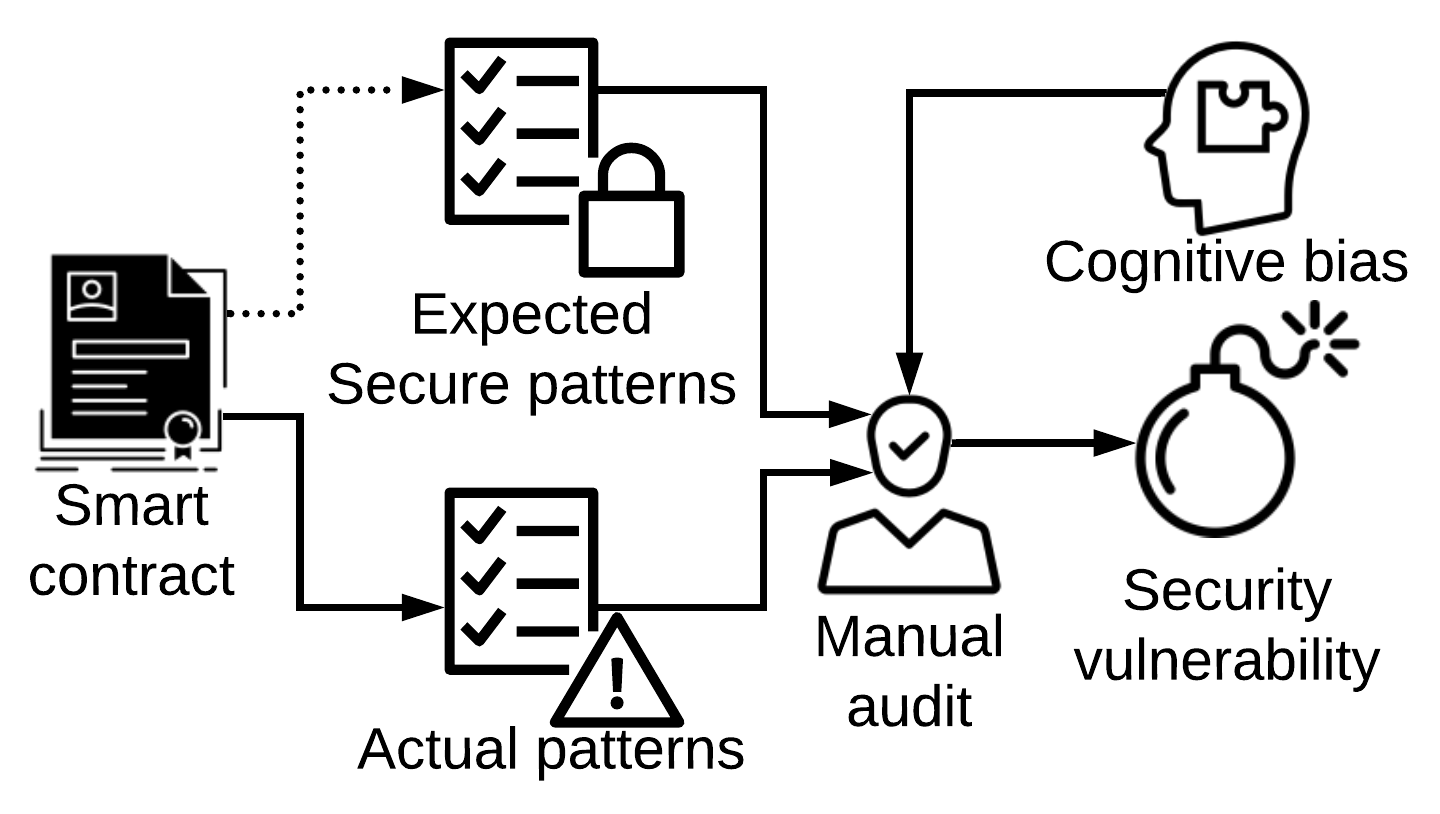}
    \caption{The workflow of cognitive bias in smart contracts.}
    \label{fig:cogn-overview}
\end{figure}

\section{Social Engineering Attacks in Ethereum}

In this work, we develop six novel social engineering attacks that target the general population of smart contract users. We subdivide these attacks into two groups: three Address Manipulation Attacks, which take advantage of specific features of Ethereum account addresses, and three Homograph Attacks, which take advantage of the fact that certain symbols look identically, but have different codes, which might affect the execution of the code. In each attack, the owner of the malicious smart contract is aiming to freeze the victim’s funds, and subsequently acquire this money through contract self-destruction, repeated withdrawals of dividends, or similar methods. In summary, our taxonomy of social engineering attacks in Ethereum smart contracts is as follows:

\begin{enumerate}
        \item \textbf{Address Manipulation Attacks} --- Ethereum public addresses.
        \begin{itemize}
            \item \attackI: Replace EOA with a non-payable contract address to incur transfer failure and revert transaction;
            \item \attackII: Pre-calculate a future contract address and replace EOA with a non-payable contract at this address;
            \item \attackIII: Exploit EVM's EIP-55 checksum insensitivity in address comparison;
        \end{itemize}
        
        \item \textbf{Homograph Attacks} --- identically looking symbols.
        \begin{itemize}
            \item \attackIV: Use dynamically-injected homograph string in a branching condition;
            \item \attackV: Replace inter-contract call (ICC) header with identically looking one to call a non-existing function;
            \item \attackVI: Suppress EVM exception by mining a function that matches a tampered ICC header.
        \end{itemize}    
    \end{enumerate}

Next, let us examine of the the social engineering attacks in detail.

\section{Attack Examples}

In this section, we demonstrate two example of our social engineering attacks. For details of the full inventory of attacks, please refer to~\cite{ivanov2021targeting}.

\subsection{Attack \attackII}


\begin{figure}
    \centering
    \includegraphics[width=\linewidth]{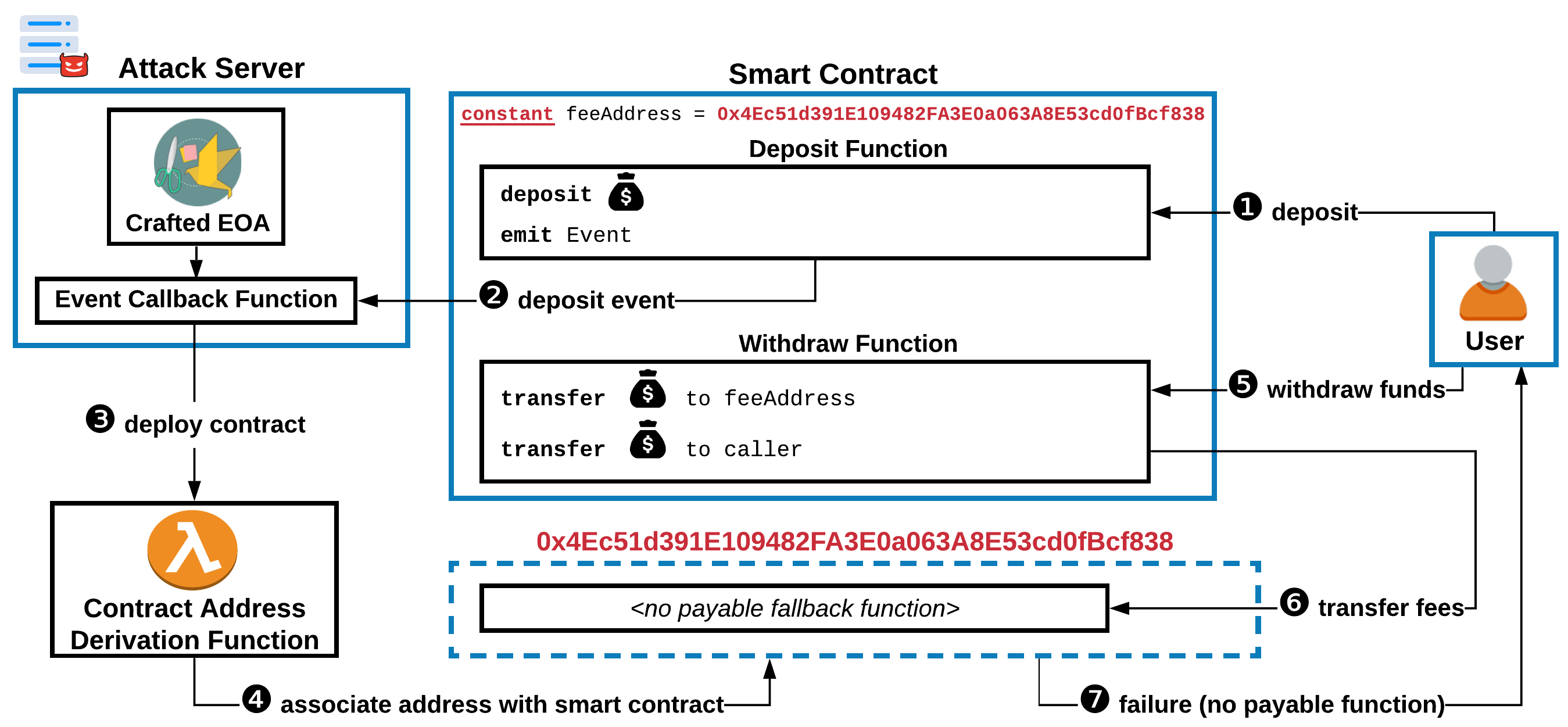}
    \caption{Workflow of attack \attackII.}
    \label{fig:a2}
\end{figure}

Consider the social engineering attack shown in Fig.~\ref{fig:a2}, in which the attacker uses a little-known contract address derivation function to pre-calculate the address of a future smart contract. Then, the attacker deploys the main smart contracts, which implements a financial scheme and charges a small service fee. The attacker hard-codes the pre-calculated address in the main smart contract. At this stage, the address points to an externally owned account, and if an auditor deploys the replica of this contract on a testnet, it will work impeccably as advertised. However, when the user deposits money to the production deployment of the contract, a server deployed by the attacker and subscribed to the deposit event, will immediately deploy a non-payable smart contract at the pre-calculated address. After some time, when the user decides to withdraw the deposited money with possible dividends, the transaction will suddenly fail because of inability to transfer the service fee to the address that is unable to accept Ether.

\subsection{Attack \attackVI}

In Ethereum, each function is encoded as a function selector, which is a 32-bit prefix of the hash of the function header, i.e., 

$$S_f = P_{32}(H_{k}(\overbrace{\text{``}f(\alpha_1,...,\alpha_n)\text{''}}^{\text{\normalsize \shortstack{function header string}}})),$$
where $P_{32}$ is a 32-bit prefix, $H_{k}$ is the \mbox{Keccak256} hash function, $f$ is the function name, and $\alpha_1,...,\alpha_n$ is the list of argument types ($0 \le n \le 16$). For example, the selector of the function $foo$ with a single 256-bit unsigned integer argument is $P_{32}(H_{k}(``foo(uint256)"))=\mathrm{0x2fbebd38}$.

In this social engineering attack example, the attacker substitutes some symbol in the header of function \texttt{foo()}, and mines a function, whose selector corresponds to this homograph header. As a result, the smart contract unpredictably calls a wrong function of another smart contract, resulting in an unpredictable behavior.



    \vspace{20pt}

    \begin{figure}
        \centering
        \includegraphics[width=\linewidth]{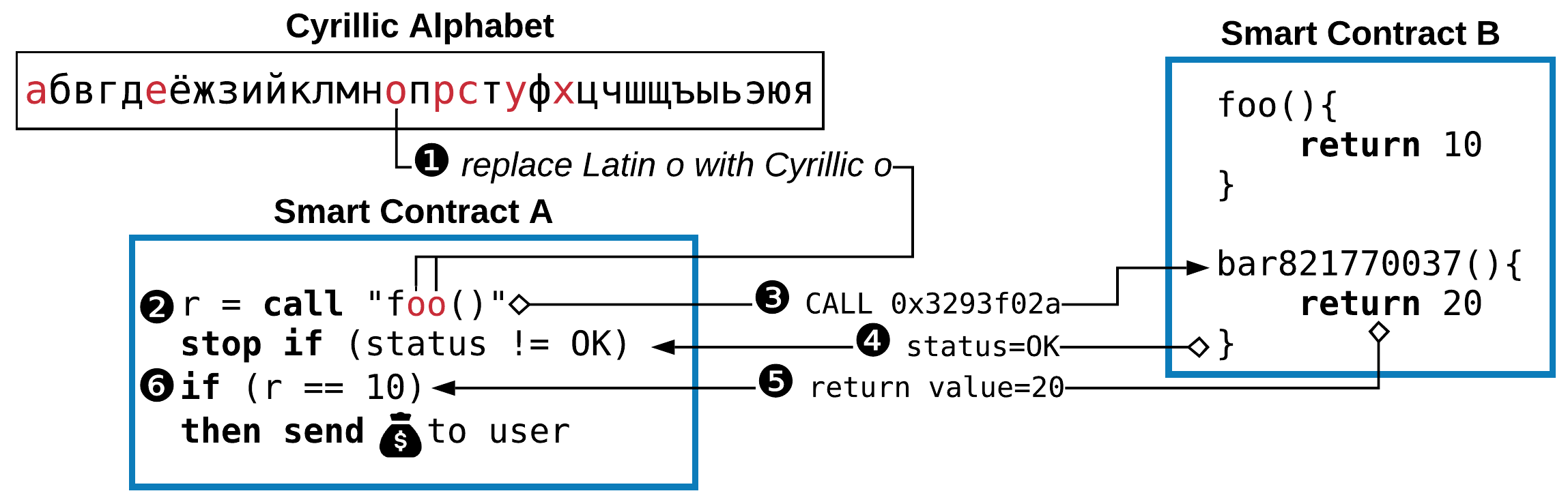}
        \caption{Workflow of attack \attackVI.}
        \label{fig:a6workflow}
    \end{figure}
    
\section{Conclusion}
In summary, we are the first to systematically study social engineering attacks in Ethereum smart contracts. We developed six zero-day attacks, and demonstrated that most of our attacks do not reveal themselves on the testnet; however, they activate their malicious capacity only at the production stage. To show that the attacks are very practical, we successfully embed their patterns in source codes of prominent existing contracts without disrupting their functionality or introducing any suspicious logic.


\end{document}